\documentclass[aip,reprint]{revtex4-1}
\usepackage{graphicx,bm,color}
\graphicspath{{./fig/}{./png/}}

\newcommand{\bra}[1]{\langle #1\rangle}
\newcommand{\nab}{\mbox{\boldmath $\nabla$} {}}
\newcommand{\ii}{{\rm{i}}}
\newcommand{\dd}{{\rm{d}}}
\newcommand{\DD}{{\rm{D}}}

\newcommand{\Rey}{{\rm{Re}}}
\newcommand{\Prz}{{\rm{Pr}_0}}
\newcommand{\Ma}{{\rm{Ma}}}

\newcommand{\nabad}{\nabla_{\rm ad}}

\newcommand{\SSSS}{\mbox{\boldmath ${\sf S}$} {}}

\newcommand{\FF}{\bm{F}}
\newcommand{\ff}{\bm{f}}
\newcommand{\xx}{\bm{x}}

\newcommand{\uu}{\bm{u}}
\newcommand{\kk}{\bm{k}}

\def\chitz{\chi_{\rm t0}}
\def\chit{\chi_{\rm t}}
\def\kappat{\kappa_{\rm t}}
\def\cgam{c_\gamma}
\def\csz{c_{\rm s0}}

\def\cp{c_{\rm p}}
\def\cv{c_{\rm v}}
\def\kf{k_{\rm f}}

\def\urms{u_{\rm rms}}

\def\etat{\eta_{\rm t}}

\def\EK{E_{\rm K}}

\newcommand{\kms}{\,{\rm km\,s}^{-1}}
\newcommand{\cm}{\,{\rm cm}}
\newcommand{\km}{\,{\rm km}}
\newcommand{\Mm}{\,{\rm Mm}}
\newcommand{\g}{\,{\rm g}}
\newcommand{\K}{\,{\rm K}}
\newcommand{\s}{\,{\rm s}}
\newcommand{\ks}{\,{\rm ks}}

\newcommand{\Sec}[1]{Sect.~\ref{#1}}

\newcommand{\Fig}[1]{Fig.~\ref{#1}}

\newcommand{\Tab}[1]{Table~\ref{#1}}

\newcommand{\yproc}[7]{, ``#4,'' In {\em #5} (ed.\ #6), pp.\ #2--#3.\ #7 (#1).}
\newcommand{\ybook}[3]{ {\em #2}.\ #3 (#1).}

\newcommand{\ypasj}[5]{, #5, {\rm Publ. Astron. Soc. Jap.\ }{\bf #2}, #3-#4 (#1).}

\newcommand{\yana}[5]{, #5, {\rm Astron.\ Astrophys.\ }{\bf #2}, #3--#4 (#1).}

\newcommand{\ynjp}[4]{, #4, {\rm New J.\ Phys.\ }{\bf #2}, #3 (#1).}
\newcommand{\yanaN}[4]{, #4, {\rm Astron.\ Astrophys.\ }{\bf #2}, #3 (#1).}

\newcommand{\smn}[3]{, #3, {\rm Month. Not. Roy.\ Astron.\ Soc.}, submitted, arXiv:#2 (#1).}
\newcommand{\ymn}[5]{, #5, {\rm Month. Not. Roy.\ Astron.\ Soc.\ }{\bf #2}, #3--#4 (#1).}

\newcommand{\yjfmN}[4]{, #4, {\rm J.\ Fluid Mech.\ }{\bf #2}, #3 (#1).}

\newcommand{\ypreN}[4]{, #4, {\rm Phys.\ Rev.\ E }{\bf #2}, #3 (#1).}

\newcommand{\yjcp}[5]{, #5, {\rm J.\ Comp.\ Phys.\ }{\bf #2}, #3--#4 (#1).}

\newcommand{\yjppN}[4]{, #4, {\rm J.\ Plasma Phys.\ }{\bf #2}, #3 (#1).}

\newcommand{\sprl}[3]{, #3, {\rm Phys.\ Rev.\ Lett.}, submitted, arXiv:#2 (#1).}

\newcommand{\papj}[3]{, #3, {\rm Astrophys.\ J.}, in press, arXiv:#2 (#1).}

\newcommand{\yapj}[5]{, #5, {\rm Astrophys.\ J.\ }{\bf #2}, #3--#4 (#1).}
\newcommand{\yapjN}[4]{, #4, {\rm Astrophys.\ J.\ }{\bf #2}, #3 (#1).}

\newcommand{\ypf}[5]{, #5, {\rm Phys.\ Fluids }{\bf #2}, #3--#4 (#1).}

\newcommand{\ygafd}[5]{, #5, {\rm Geophys.\ Astrophys.\ Fluid Dynam. }{\bf #2}, #3--#4 (#1).}

\newcommand{\yjour}[6]{, #6, {\rm #2} {\bf #3}, #4--#5 (#1).}
\newcommand{\yjourN}[5]{, #5, {\rm #2} {\bf #3}, #4 (#1).}

\newcommand{\EQ}{\begin{equation}}
\newcommand{\EN}{\end{equation}}
\newcommand{\ba}{\begin{eqnarray}}
\newcommand{\ea}{\end{eqnarray}}

\newcommand{\Eq}[1]{Eq.~(\ref{#1})}
\newcommand{\eq}[1]{(\ref{#1})}
\newcommand{\Eqs}[2]{Equations~(\ref{#1}) and (\ref{#2})}
\newcommand{\EEqs}[2]{Equations~(\ref{#1})--(\ref{#2})}

\def\nnn{\bm{\hat n}}
\def\eee{\bm{\hat e}}
\def\sigmaSB{\sigma_{\rm SB}}

\def\ii{{\rm i}}

\graphicspath{{./fig/}{./png/}}

\begin{document}

\preprint{NORDITA-2020-100}

\title{Turbulent radiative diffusion and turbulent Newtonian cooling}

\author{Axel Brandenburg}
\email[]{brandenb@nordita.org}
\homepage[]{https://www.nordita.org/~brandenb/}
\affiliation{Nordita, KTH Royal Institute of Technology and Stockholm University,
Hannes Alfv\'ens v\"ag 12, 10691 Stockholm, Sweden}
\affiliation{Department of Astronomy, Stockholm University, 10691 Stockholm, Sweden}
\affiliation{McWilliams Center for Cosmology and Department of Physics, Carnegie Mellon University, 5000 Forbes Ave, Pittsburgh, PA 15213, USA}
\affiliation{School of Natural Sciences and Medicine, Ilia State University, 3-5 Cholokashvili Avenue, 0194 Tbilisi, Georgia}

\author{Upasana Das}
\affiliation{Nordita, KTH Royal Institute of Technology and Stockholm University,
Hannes Alfv\'ens v\"ag 12, 10691 Stockholm, Sweden}

\date{\today,~ $ $Revision: 1.54 $ $}

\begin{abstract}
Radiation transport plays important roles in stellar atmospheres,
but the effects of turbulence are being obscured by other effects
such as stratification.
Using radiative hydrodynamic simulations of forced turbulence,
we determine the decay rates of sinusoidal large-scale temperature
perturbations of different wavenumbers in the optically thick and thin
regimes.
Increasing the wavenumber increases the rate of decay in both regimes,
but this effect is much weaker than for the usual turbulent diffusion
of passive scalars, where the increase is quadratic for small wavenumbers.
The turbulent decay is well described by an enhanced Newtonian cooling
process in the optically thin limit, which is found to show a weak
increase proportional to the square root of the wavenumber.
In the optically thick limit, the increase in turbulent decay is somewhat
steeper for wavenumbers below the energy-carrying wavenumber of the
turbulence, but levels off toward larger wavenumbers.
In the presence of turbulence, the typical cooling time is comparable
to the turbulent turnover time.
We observe that the temperature takes a long time to reach equilibrium
in both the optically thin and thick cases, but in the former, the
temperature retains smaller scale structures for longer.
\end{abstract}


\maketitle

\section{Introduction}

An important property of turbulence is the mixing of fields that are
advected by the flow.
The simplest example is that of a passive scalar, a quantity that does
not backreact on the flow.
The magnetic field is another popular example, because for weak field
strengths, it can be treated as a passive vector field, making the
mathematics more straightforward compared to the fully nonlinear case.
Even the flow itself is mixed by the turbulence, which is a much harder
problem.
This leads to turbulent viscosity, which acts as an enhanced molecular
viscosity, although there can be additional important effects if the
turbulence is anisotropic.
Examples of additional effects occur in stratified flows in the presence
of rotation.
Such flows can become differentially rotating through what is called
the $\Lambda$ effect \cite{Rue80}.
It is a nondiffusive effect, analogous to the $\alpha$ effect in
mean-field dynamo theory \cite{Mof78,KR80}.
These nondiffusive effects have led to significant attention in
astrophysics.
Scalars, active or passive, have received comparatively less attention,
because nondiffusive effects are generally less profound, but see R\"adler
et al. \cite{Rae11} for the slow-down of turbulent diffusion in certain
compressive flows.

Prandtl suggested that turbulence has a smoothing effect---just like
molecular diffusion.
The molecular diffusion coefficient is generally proportional to the
product of the typical velocity of the molecules, which is essentially
the sound speed, and the typical mean-free path between collisions.
Prandtl generalized this to turbulence by using the product of the
typical velocity of the turbulent eddies and their correlation length,
which he referred to as the mixing length.
Important applications of turbulent mixing in astrophysics include
turbulent convection in the Sun and stars, as well as mixing of chemicals
in the Galaxy.
The latter is a typical case of a passive scalar, while in the former case,
the quantity that is being mixed is the specific entropy, which is an
active scalar, because it affects the density in the momentum equation
and can lead to buoyancy.
Furthermore, the resulting turbulent diffusion is an enhancement not of
molecular diffusion, but of photon diffusion, which is also referred to
as radiative diffusion.

Radiative diffusion comes in two different forms: optically thick and
optically thin.
Optically thick is the usual case, where the mean-free path of photons
is short compared to the typical scales of the flow.
Optically thin, by contrast, means that the photons can propagate over
large distances before they are absorbed and re-emitted again.
Radiative diffusion ceases to exist in this case and we have to deal
instead with an essentially nonlocal process.
The effect of radiation now decreases with increasing mean-free path of
the photons, contrary to the diffusive case where it increases.
The relevant process, in this case, is Newtonian cooling \cite{Schatzman},
where the cooling rate is directly proportional to the radiation energy
density rather than the divergence of its gradient.
The cooling timescale is the {\em ratio}
of mean-free path to some relevant photon speed, rather than turbulent
diffusion, whose coefficient is proportional to the {\em product} of
mean-free path and the relevant photon speed.

In astrophysics, one usually thinks of optically thin processes being
those that happen above the photosphere of a star, where photons can
travel all the way to infinity.
However, even below the photosphere, a process can be optically thin if
we look at small length scales, because then the photon mean-free path
again exceeds the relevant scale of the flow structures.
In this paper, we are interested in the effects of turbulence, especially
in this optically thin limit.

There is actually a curious analogy between the optically thin limit,
where cooling becomes less efficient at small length scales \cite{BD20},
and turbulent diffusion, which also becomes less efficient at small
length scales\cite{BRS08,BSV09}, although this concerns here the length
scales of the mean fields.
This is because turbulent diffusivity is not just a coefficient, but an
integral kernel in a convolution with the mean temperature\cite{Rae76}.
The Fourier transformation of this kernel falls off with wavenumber
approximately like a Lorentzian, which is analogous to the case of
radiative transfer in the optically thin case \cite{US66}.
We may therefore ask: how does the combined effect of turbulence and
small optical thickness modify turbulent diffusion at small length scales?

We begin by reviewing some basics about the cooling time as a function
of mean free path in \Sec{CoolingCurve}, following which we describe our
numerical simulations in \Sec{TurbulenceSim}, and finally present our
results in \Sec{Results}.
Our conclusions and scope for future work are given in \Sec{Conclusions}.

\section{The cooling curve}
\label{CoolingCurve}

In compressible hydrodynamics, the energy equation can be written
in terms of the specific entropy $s(\xx,t)$ as
\EQ
\rho T{\DD s\over\DD t}=-\nab\bm\cdot\FF_{\rm rad}+2\rho\nu\SSSS^2,
\label{DsDt}
\EN
where $\rho$ is the density, $T$ the temperature, 
$\DD/\DD t=\partial/\partial t+\uu\bm\cdot \nab$ the advective derivative,
$\FF_{\rm rad}$ the radiative flux, $\nu$ the viscosity,
and $\SSSS$ the traceless rate-of-strain tensor with the components
${\sf S}_{ij}=(\partial_i u_j+\partial_j u_i)/2-\delta_{ij}\nab\bm\cdot\uu/3$.
The negative divergence of $\FF_{\rm rad}$ is calculated as the imbalance
of the intensity and the source function integrated over all frequencies
$\tilde\nu$ and all directions\cite{Nor82}, i.e.,
\EQ
-\nab\bm\cdot\FF_{\rm rad}=\int_0^\infty\kappa_\nu\rho
\oint_{4\pi}(I_{\tilde\nu}-S_{\tilde\nu}) \,\dd\Omega\,\dd\tilde\nu,
\label{fff}
\EN
where $\kappa_{\tilde\nu}$ is the opacity per unit mass,
$I_{\tilde\nu}(\xx,t,\nnn)$ is the specific intensity corresponding
to the energy that is carried by radiation per unit area, per unit
time, in the direction $\nnn$, per unit solid angle $\dd\Omega$, and
$S_{\tilde\nu}(\xx,t)$ is the source function.
Throughout this work, we make the gray approximation and thus work
with frequency-integrated quantities, which amounts to dropping the
subscript $\tilde\nu$.
In the gray approximation, $I(\xx,t,\nnn)$ obeys the
radiative transfer equation,
\EQ
\nnn\bm\cdot\nab I=-\kappa\rho\, (I-S),
\label{RT-eq}
\EN
which is solved along a set of rays in different directions $\nnn$
using the method of long characteristics \cite{Rijkhorst}.
The source function is here written as $S=(\sigmaSB/\pi)\,T^4$,
where $\sigmaSB$ is the Stefan--Boltzmann constant.
The photon mean-free path is $\ell=(\kappa\rho)^{-1}$ and $\kappa$
is a suitably averaged ``gray'' opacity.

By assuming infinitesimally small temperature perturbations, Spiegel
\cite{Spi57} linearized \Eqs{DsDt}{RT-eq} and found that for perturbations
of the form proportional to $\exp(\ii\kk\bm\cdot\xx+\lambda t)$,
the inverse relaxation (or cooling) time $\lambda$ of the temperature
perturbations, or decay rate, is given by
\begin{equation}
\lambda={c_\gamma\over\ell}\,
\left(1-\frac{\arctan k\ell}{k\ell} \right),
\label{S57expr}
\end{equation}
where $k=|\kk|$ is the wavenumber,
\begin{equation}
c_\gamma=16\sigmaSB T^3/\rho \cp
\label{cgam_def}
\end{equation}
is the characteristic velocity of photon diffusion \cite{BB14},
and $\cp$ the specific heat at constant pressure.
The dependence $\lambda(k\ell)$ is what we call in this paper the
cooling curve.
A useful form of the above expression can be obtained under the
Eddington approximation \cite{Edw90}, where one expands \Eq{RT-eq}
in terms of moments of $\nnn$ under the closure assumption that
$\oint\hat{n}_i\hat{n}_j I\,\dd\Omega =\frac{1}{3}\delta_{ij}\oint
I\,\dd\Omega$.
This yields a closed equation $\frac{1}{3}(\ell\nabla)^2J=J-S$
for the mean intensity $J=\oint I\,\dd\Omega/4\pi$.
The cooling rate is then
\begin{equation}
\lambda\approx{c_\gamma\over\ell}\,{k^2\ell^2/3\over1+k^2\ell^2/3}.
\label{US66expr}
\end{equation}
It is convenient to introduce now the radiative diffusivity,
$\chi=c_\gamma\ell/3$, which also clarifies why $c_\gamma$ is called
the characteristic velocity of photon diffusion.

\begin{figure}
\includegraphics[width=\columnwidth]{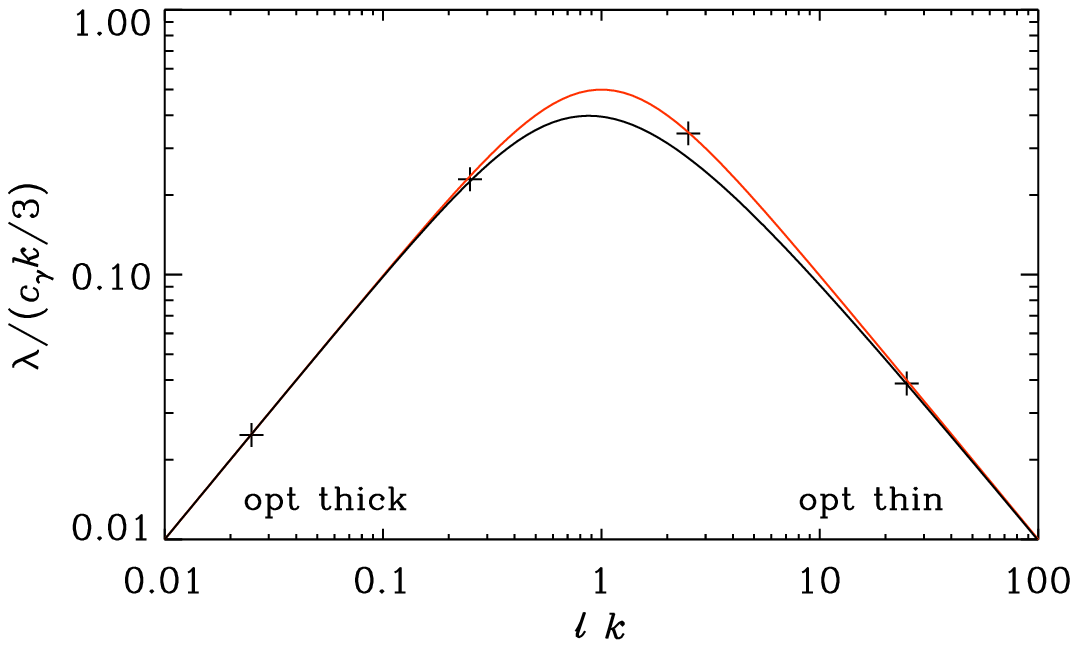}
\caption{
The analytic cooling curve (black), the cooling curve under the Eddington
approximation (red), and the results from numerical simulations
(plus signs).
\label{pdecay_comp0}}
\end{figure}

\EEqs{S57expr}{US66expr} apply to the case of three-dimensional
(3-D) variations of the temperature.
In our 3-D numerical experiments, however, we restrict ourselves to
one-dimensional (1-D) variations of the {\em mean} temperature profile.
In that case, the relevant version of \Eq{US66expr} becomes \cite{BB14}
\begin{equation}
\lambda\approx{c_\gamma\over\ell}\,{k^2\ell^2/3\over1+k^2\ell^2}
\quad\mbox{(1-D perturbations)}.
\label{US66expr1D}
\end{equation}
The corresponding version of \Eq{S57expr} then takes the form
\begin{equation}
\lambda={c_\gamma\over\ell}\,
\left(1-\frac{\arctan \sqrt{3} k\ell}{\sqrt{3} k\ell} \right)
\quad\mbox{(1-D perturbations)}.
\label{S57expr1D}
\end{equation}
In \Fig{pdecay_comp0}, we compare $\lambda(k\ell)$ obtained from the
exact equation (red curve) with the approximate $\lambda(k\ell)$
obtained under the Eddington approximation (black curve) for the
relevant 1-D \Eqs{US66expr1D}{S57expr1D}.
Our numerical solution for $\lambda$, which is based on only six rays,
depends on the choice of weight factors used in the angular integration.
The weight factors have been chosen such that our numerical results
(plus signs) agree with the Eddington approximated solution \cite{BB14}.
The basic question we want to answer is how the cooling curve gets
modified in the presence of turbulence.
We expect the effective $\lambda$ to be enhanced, at least in the
optically thick limit, where $k\ell\ll1$; however, we do not know what
to expect in the optically thin case, where $k\ell\gg1$, and how it
depends on the scale of the turbulent eddies.
To address these questions, we now perform turbulence simulations.
We are particularly interested in the regime of moderate temperatures,
where the radiation pressure can be ignored.

\section{Turbulence simulations}
\label{TurbulenceSim}

\subsection{Comments on astrophysical conditions}
\label{Comment}

The conditions in the Sun are extremely inhomogeneous owing to
tremendous stratification.
The density changes by nearly six orders of magnitude across the
convection zone and the temperature by a factor of about 300.
This fact alone can introduce new phenomena such as the spontaneous
formation of magnetic flux concentrations \cite{BRK16,PB18}.
At a more elementary level, the addition of gravity leads to convection
and thereby to turbulent motions, which have been the subject of
numerous simulations for a long time \cite{Nor82,SN98}.
Newtonian cooling also plays important roles in the atmospheres of planets
\cite{ZJBYAM21}, where turbulence is not always explicitly invoked and
therefore the role of turbulence needs to be parameterized \cite{SPRS21}.

In the Sun, the microphysical viscosity is about twelve orders of
magnitude smaller than the estimated turbulent viscosity, and over
eight orders of magnitude smaller than the radiative diffusivity near
the surface.
Numerical simulations have therefore routinely employed numerical tools
that allow the simulations to proceed by dissipating sufficient energy
in local regions where necessary.
This precluded the study of turbulent Newtonian cooling,
because the small-scale turbulent motions have already been altered by
such numerical modeling \cite{Leenaarts20}.
An additional complication is partial ionization, which tends to make
the transition from the deeper optically thick layers to the surface
very sharp in a stratified system.
However, one could imagine it to introduce new effects of its own if we
arranged the average temperature of the domain such that it lies exactly
in the middle between those of a fully ionized and a neutral medium,
as has been done in some other recent experiments \cite{BB16}.
In those cases, however, Newtonian cooling does not necessarily play
any obvious role.

To accomplish our goal of identifying turbulent effects in optically thin
and thick turbulent flows, we resort to the study of a minimal system
where isotropic homogeneous turbulence is produced by a stirring force
instead of convection.
The viscosity is kept constant, but we consider different values to
assess the dependence of our results on the Reynolds number.
Partial ionization effects are ignored and other complications
from adopting a realistic equation of state are not included.
We also restrict our attention to the study of vortical forcing.
The concept of turbulent mixing is likely to be similar also for
irrotational forcing, but other poorly understood features of such
turbulence such as a pileup of kinetic energy near the dissipative
subrange (bottleneck effect) are known to occur in such cases \cite{MB06}.
It may play a role in interstellar turbulence \cite{Fed10}, which
motivates a more extended future study of turbulent Newtonian cooling
for irrotational forcing.

\subsection{Basic equations and thermodynamic relationships}
\label{BasicEquations}

For the purpose of our present study, we restrict ourselves to studying
a turbulent flow in a triply periodic domain of size $L^3$ by applying
plane wave forcing throughout the domain.
We therefore solve the equations,
\EQ
\rho{\DD\uu\over\DD t}=-\nab p + \rho\ff +\nab\bm\cdot(2\rho\nu\SSSS),
\label{DuDt}
\EN
\EQ
{\DD\ln\rho\over\DD t}=-\nab\bm\cdot\uu,
\label{dlnrho}
\EN
where $p$ is the pressure, $\uu$ the velocity, and $\ff$ the forcing function.
In \Eq{DuDt}, we have ignored the radiation force $(\rho\kappa/c)\,\FF_{\rm rad}$,
where $c$ is the speed of light, as mentioned above.
This term is unimportant for the temperatures considered in this work.
Nevertheless, the coupled set of equations \eq{DsDt}, \eq{DuDt}, and
\eq{dlnrho} makes $s(\xx,t)$ an active scalar, because it is related to
$p$ and $\rho$ through
\EQ
s=\cv\ln p-\cp\ln\rho+s_0,
\label{seqn}
\EN
where $\cv$ is the specific heat at constant volume
and $s_0$ is an irrelevant constant.
This equation follows from the first law of thermodynamics, written in the
form $T\dd s=\dd e+p\dd(\rho^{-1})$, where $e=\cv T$ is the internal energy
for a perfect gas, and the ideal gas equation relating the temperature
to $p$ and $\rho$ through
\EQ
(\cp-\cv)\,T=p/\rho.
\EN
We then have $\dd s=\cv\,\dd\ln T-(\cp-\cv)\,\dd\ln\rho$.
Using the differentiated ideal gas equation,
$\dd\ln T+\dd\ln\rho=\dd\ln p$, we arrive at
\Eq{seqn} after integration.
For the ratio of specific heats, $\gamma=\cp/\cv$, we assume $\gamma=5/3$,
which is appropriate for a monatomic gas such as fully ionized hydrogen
at the temperatures considered here ($T\approx40,000\K$).

In our numerical work, we use dimensionful quantities, where length is
measured in megameters (Mm), speed in $\kms$, and temperature in kelvin.
We also use the symbol $\nabad=1-1/\gamma=0.4$, which is the adiabatic
value of what is in astrophysics commonly referred to as the
double logarithmic temperature gradient, $\nabla\equiv\dd\ln T/\dd\ln p$
\cite{KW90}.
Using this, $\cp$ can then be written as $\cp={\cal R}/(\mu\nabad)$,
where we have used $\cp-\cv={\cal R}/\mu$, with
${\cal R}=8.314\times10^7\cm^2\s^{-2}\K^{-1}$ being the universal gas constant
and $\mu=0.6$ the mean molecular weight.
We then find $\cp=0.035\km^2\s^{-2}\K^{-1}$.

\subsection{Turbulent forcing}
\label{TurbulentForcing}

To simulate a turbulent flow, we apply nearly monochromatic
forcing with an average forcing wavenumber $\kf$.
The forcing function changes abruptly from one time step to the next,
i.e., $\bra{\ff(\xx,t)\ff(\xx,t')}$ is proportional to $\delta(t-t')$,
where $\delta$ is the Dirac $\delta$ function.
The forcing is then said to be $\delta$ correlated in time.
The smallest wavenumber in the cubic domain of side length $L$ is $k_1=2\pi/L$.
The ratio $\kf/k_1$ is the scale separation ratio, for which we consider
the values 1.5 and 10.

For the forcing function $\ff$, we select randomly at each time step a
phase $-\pi<\varphi\le\pi$ and the components of the wavevector $\kk$
from many possible discrete wavevectors with lengths in a certain range
around a given value $\kf$.
In this way, the adopted forcing function
\begin{equation}
\ff(\xx,t)={\rm Re}\{{\cal N}\tilde{\ff}(\kk,t)\exp[i\kk\cdot\xx+i\varphi]\}
\end{equation}
is white noise in time and consists of plane waves with average wavenumber
$\kf$.
Here, $\xx$ is the position vector and
${\cal N}=f_0(\csz\kf\delta t)^{1/2}$
is a normalization factor, where $\delta t$ is the time step and $f_0$
is an amplitude factor.
In this formulation, the averaged forcing is independent of $\delta t$.
To ensure that $\tilde{\ff}$ is solenoidal, i.e., perpendicular
to $\kk$, we write is as
\EQ
\tilde{\ff}({\kk})=(\kk\times\eee)/[\kk^2-(\kk\cdot\eee)^2]^{1/2},
\EN
where $\eee$ is an arbitrary unit vector that are not aligned with $\kk$.
Note that $|\ff|^2=1\km^4\s^{-4}\Mm^{-2}$.
The coefficient $f_0$ is chosen such that the velocity is about 10\%
of the sound speed.

\subsection{Initial temperature profile and parameters}

We adopt a sinusoidal temperature perturbation and write $T(x,t=0)$
in the form
\EQ
T(x,t=0)\,=\,T_0+T_1\sin kx,
\EN
where $k=k_1=1\Mm^{-1}$ is chosen.
This implies that $L=2\pi\Mm\approx6.28\Mm$.
We choose $\csz=30\kms$, so that $T_0\approx40,000\K$ and $c_\gamma=3.9\kms$.
The temperature perturbation is taken to be $T_1=2000\K$, and
periodic boundary conditions are assumed for all quantities.

We define the Mach and Reynolds numbers as
\EQ
\Ma=\urms/\csz,\quad\Rey=\urms/\nu\kf.
\EN
For a forcing amplitude $f_0=0.01\km^2\s^{-2}\Mm^{-1}$, we have
$\urms\approx2.2\kms$, so that $\Ma\approx0.08$.
Using $\nu=10^{-3}\Mm\km\s^{-1}$ and $\kf=10\Mm^{-1}$, we have
$\Rey\approx230$, while for $\kf=1.5\Mm^{-1}$, we have $\Rey\approx1200$.
To determine the microphysical Prandtl number in the optically thick
regime, we define
\EQ
\Prz=3\nu k_1/\cgam,
\EN
so that the Prandtl number is $\nu/\chi=\Prz/(k_1\ell)$.
Here we have used $\chi=c_\gamma\ell/3$ for the radiative diffusivity.
Following earlier work \cite{BB14}, we choose for the mean density
the value $\rho_0=4\times10^{-4}\g\cm^{-3}$.

We determine the effective $\lambda$ from the time evolution of the
decay of the sinusoidal perturbation, which we monitor by taking the
difference between the maximum and minimum temperatures at each time.
This turns out to be reasonably accurate and we use a time interval
during which the decay is exponential.

\subsection{Numerical technique}

We perform numerical simulations with the {\sc Pencil Code}
(\url{https://github.com/pencil-code}), which is a public MHD code that
is particularly well suited for simulating turbulence \cite{PencilCollab}.
We solve Eqs.~(\ref{DsDt}), (\ref{DuDt}), and (\ref{dlnrho}) with
sixth-order finite differences \cite{Bra03}.
Equation~(\ref{RT-eq}) is solved with second-order accurate finite
differences along the coordinate directions and the diagonals, i.e.,
altogether 26 rays.
The radiation transport has been parallelized in the {\sc Pencil Code}
by splitting the calculation into parts that are local and nonlocal with
respect to each processor \cite{HDNB06}.
Two parts are compute-intensive, but require no communication, and one
part is nonlocal, but does not require waiting for any computation to
be done and is therefore fast.
We use the third-order time-stepping scheme of Williamson \cite{Wil80}.

The code's local cooling and heating properties have been verified
\cite{BB14}, and its cooling time has been compared with the analytic
cooling time obtained by Spiegel \cite{Spi57}.
It is this cooling time that determines the relevant time step
constraint for radiation simulations \cite{Freytag12,BD20}, and not some
generalization of the usual Courant condition \cite{DSJ12}.
The latter would erroneously imply a limiting time step that is
proportional to the mesh spacing, when it is actually quadratic in the
mesh spacing in the optically thick limit and independent of mesh spacing
in the optically thin limit.

The code has been applied to sunspot simulations \cite{HNSS07}, and to
a range of more idealized problems of atmospheric stability \cite{BB14}
and magnetic spot formation \cite{BB16,PB18}.
It should be emphasized that our calculations classify as direct numerical
simulations in the sense that the equations are solved as stated, albeit
with unrealistically large viscosity and unrealistically small opacities
compared with solar conditions.
By comparison, most simulations of solar convection are performed using
large eddy simulations \cite{SN98,Rem09,Freytag12}.

\subsection{Comment on numerical convergence and accuracy}
\label{Convergence}

The fact that the {\sc Pencil Code} uses sixth-order finite differences
and a third-order time-stepping scheme does not tell us much about the
actual accuracy and convergence of our results.
For example, the longer a simulation, the more numerical errors should
accumulate, but this is not normally seen.
This was recently addressed in a study comparing the numerical accuracy
of turbulence and waves \cite{RPBKKM20}.
In that study, it was concluded that the existence of a forward cascade
in turbulence prevents the systematic loss of energy at small scales,
where discretization errors are the largest.
This is not the case for waves, which therefore need to be solved
with much more care.
Another such example was a recent three-dimensional study of
electromagnetic waves \cite{BS21}.
In this and the previous case, it proved advantageous to use an exact
time integration under the assumption that the turbulent source is
unchanged between two time steps.
This approximation is justified because at high wavenumbers, the
relevant timescale of waves is much shorter than that of turbulence
\cite{RPBKKM20}.

Also radiation can introduce short time scales.
As already alluded to in the beginning, the cooling time can be
very short and severely restrict the relevant time step constraint
\cite{Freytag12,BD20}.
This makes the simulations very costly \cite{DSJ12}, but we are not
aware of any reports on loss of accuracy in such cases.
Furthermore, in the present studies, we are probably not affected by
this constraint, because turbulent Newtonian cooling only plays a role
when the turbulent time scales are short compared with radiative ones.
This is here not the case.

\begin{table}[t!]\caption{
Summary of the parameters for the series of runs presented in this paper.
Within each series of runs, $\kappa$ and $\ell$ are varied.
}\vspace{12pt}\centerline{\begin{tabular}{ccccccc}
Series & $k/k_1$ & $\kf/k_1$ & $\urms/\cgam$ & $\Prz$ & $\Rey$ & $N^3$ \\
\hline
A'& 1 & 10  & 0.50 & $8\times10^{-3}$ &  20 &  $64^3$ \\
A & 1 & 10  & 0.59 & $8\times10^{-4}$ & 230 & $256^3$ \\
B & 1 & 1.5 & 0.62 & $8\times10^{-3}$ & 160 &  $64^3$ \\
C & 3 & 1.5 & 0.46 & $8\times10^{-4}$ &1200 & $256^3$ \\
D & 6 & 1.5 & 0.46 & $8\times10^{-4}$ &1200 & $256^3$ \\
\label{Tsum}\end{tabular}}\end{table}

\begin{figure*}
\includegraphics[width=.95\textwidth]{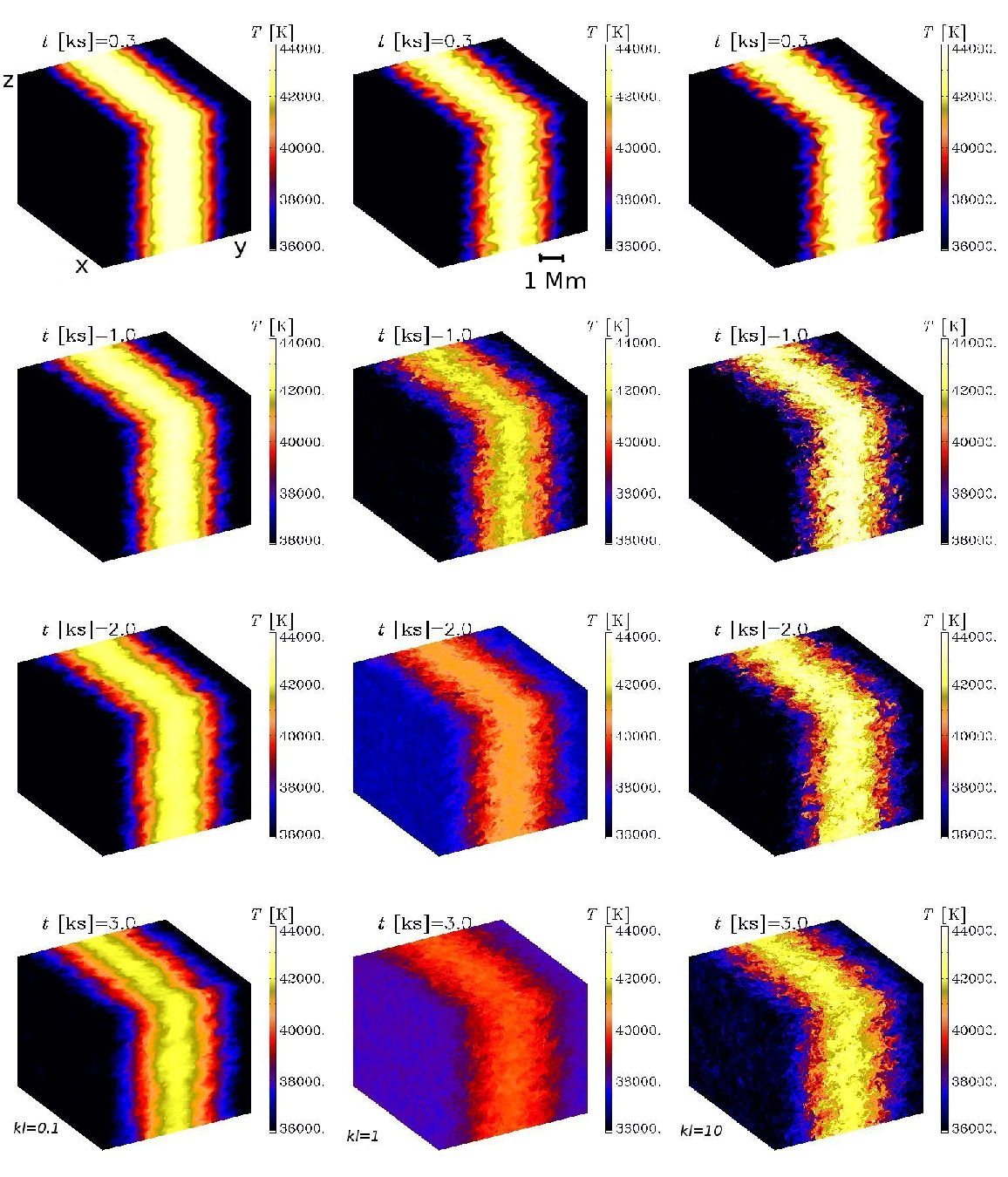}
\caption{
Temperature on the periphery of the computational domain for $k\ell=0.1$
(left column), $k\ell=1$ (middle column), and $k\ell=10$ (right column),
for $t=0.3\ks$, $t=1\ks$, $t=2\ks$, and $t=3\ks$ (top to bottom).
The $(x,y,z)$ coordinates are indicated in the first panel and the
approximate $1\Mm$ scale is shown in the second panel.
\label{ABC}}
\end{figure*}

\section{Results}
\label{Results}

\subsection{Range of simulations and qualitative aspects}
\label{Qualitative}

In this section, we present the results for $\lambda$ obtained using
various values of the forcing wavenumber $\kf$ and the wavenumber $k$
of the initial perturbation.
The Reynolds number varies between 20 and 1200, and the number of mesh
points, $N^3$, is varied between $64^3$ and $256^3$; see \Tab{Tsum}.
The values of $\Prz$ are then small, as is also expected for the Sun,
and they are $8\times10^{-3}$ for our runs with $64^3$ mesh points
(small Reynolds numbers) and $8\times10^{-4}$ for $256^3$ mesh points
(larger Reynolds numbers).
For each series of runs, we perform simulations where we vary the opacity
$\kappa$ and thereby $\ell$; see \Fig{ABC} for visualizations of $T$ on
the periphery of the computational domain for runs of Series~A for
$k\ell=0.1$, $1$, and $10$ and at different times (in kiloseconds [ks]).
We see that the large-scale temperature contrast (wavenumber
$k=k_1$) decreases the fastest for $k\ell=1$, and more slowly for
$k\ell=0.1$ and $10$.
For $k\ell=10$, however, which is the optically thin case, the temperature
retains smaller scale structures for longer.

\begin{figure}
\includegraphics[width=\columnwidth]{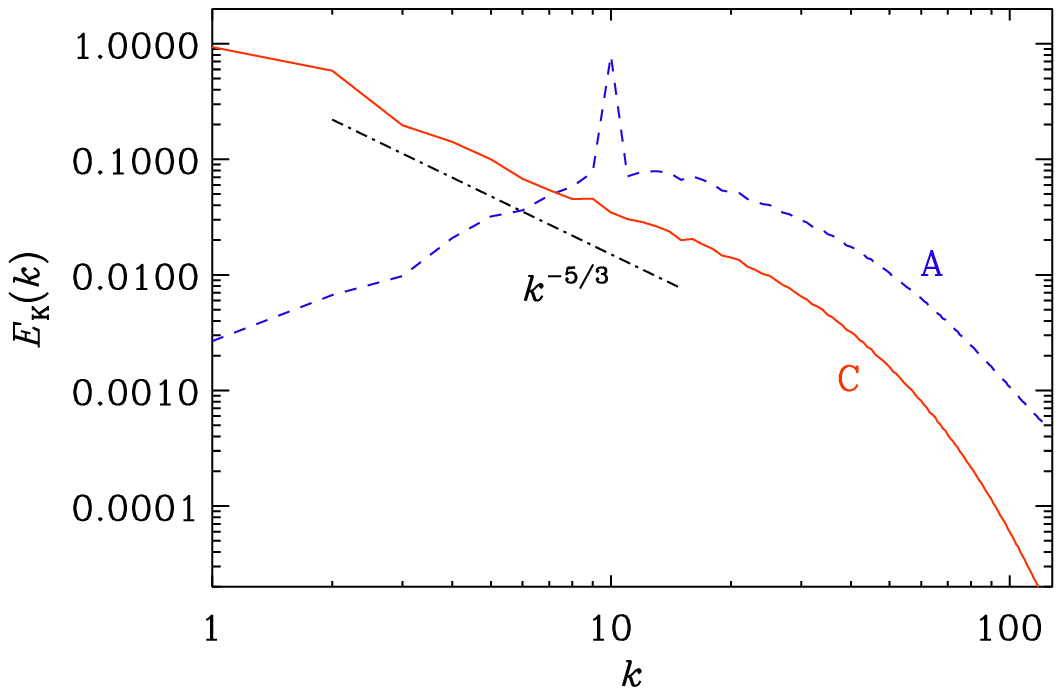}
\caption{
Kinetic energy spectra for Series~A (dashed blue) and
Series~C (solid red).
The $k^{-5/3}$ slope is overplotted for comparison.
\label{pspecm}}
\end{figure}

\subsection{Kinetic energy spectra}
\label{Kinetic}

In spite of the forcing being monochromatic, the resulting turbulence
is excited over a broad range of scales.
This is demonstrated in \Fig{pspecm}, where we plot kinetic energy
spectra, $\EK(k)=\int|\tilde{\uu}|^2k^2\dd\Omega_k$, for Series~A and C.
Here, $\tilde{\uu}$ is the Fourier transformation of $\uu$, and $\dd\Omega_k$
is the solid angle differential in wavenumber space.
The spectra are normalized such that $\int\EK(k)\,\dd k=\urms^2/2$.
In the case of Series~C, where $\Rey=1200$ and $\kf/k_1=1.5$, there is a
short inertial range $\propto k^{-5/3}$ together with a bottleneck, i.e.,
a shallower spectrum near the dissipative subrange \cite{Fal94,Zheng21}.
We note that the bottleneck effect is physical, but much weaker in the
one-dimensional spectra that are accessible to laboratory and atmospheric
measurements \cite{Dobler}.
It is also seen in the highest resolution turbulence simulations today
\cite{Fed21}.

In the simulations with larger scale separation (Series~A), however,
the spectrum is more peaked around $k=\kf$.
This occurrence of this spike at $\kf$ is partially explained by the
smaller Reynolds number ($\Rey=230$ in this case).

In general, higher scale separation allows us to see more clearly
the various mean-field effects.
In this connection, we must remember that the standard concept of
turbulent diffusion does require sufficient scale separation and that the
lack of scale separation requires one to study the full scale dependence,
in which case turbulent diffusion corresponds to an integral kernel
in real space, or a multiplication with a $k$-dependent diffusivity in
Fourier space \cite{BRS08}.
For this reason, we also discuss the aspect of scale dependence below.

\subsection{Quantitative results for turbulent cooling}
\label{TurbulentCooling}

\begin{figure}[t!]
\includegraphics[width=\columnwidth]{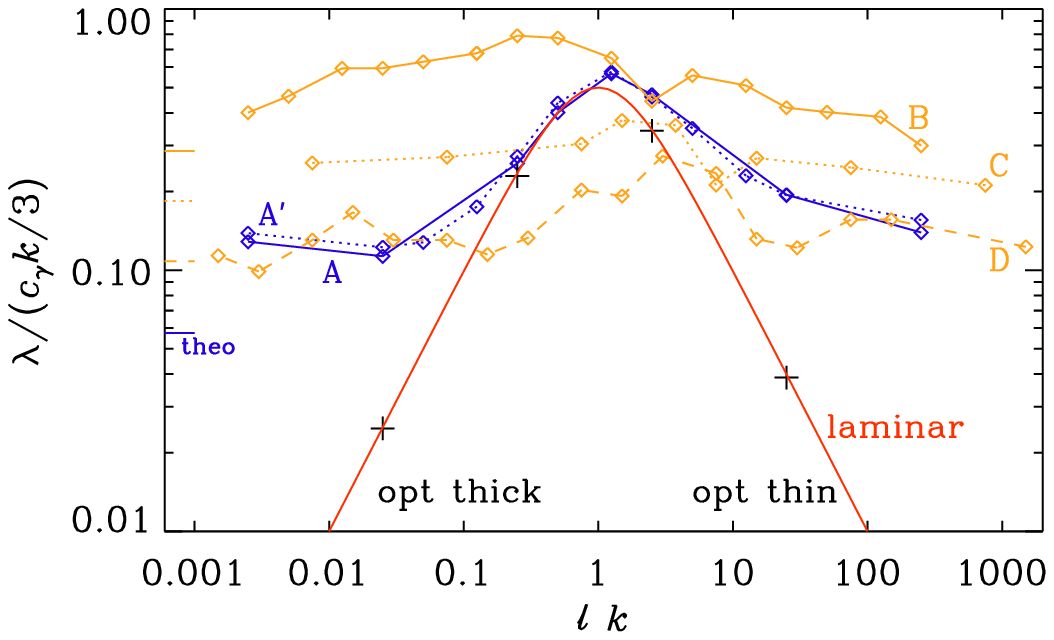}
\caption{
Effective turbulent decay rate $\lambda$ versus $k\ell$ for
Series~A' (dotted blue), A (solid blue), B (solid orange),
C (dotted orange), and D (dashed orange).
The short lines on the left $y$ axis, with line types matching those of the curves
in the plot, give the theoretical expectations explained in the text.
The curve for the laminar case is shown in red.
\label{pdecay_comp}}
\end{figure}

In \Fig{pdecay_comp}, we plot $\lambda$ versus $k\ell$ and compare with
the laminar case shown in \Fig{pdecay_comp0}.
In all the cases, we see that $\lambda$ is enhanced relative to the
laminar curve.
Varying the viscosity, and thereby changing $\Rey$ from 20 (Series~A')
to 230 (Series~A), has a very minor effect; compare the dotted and solid
blue lines for $k/\kf=0.1$ in \Fig{pdecay_comp}.
Decreasing $\kf/k_1$ from 10 to 1.5, that is, increasing $k/\kf$ from 0.1
(Series~A) to 0.7 (Series~B), has a more significant effect, and $\lambda$
is seen to increase by a factor that is between 4 and 8, depending on
the value of $k\ell$.

Keeping the value of $\kf$ unchanged and increasing $k/\kf$ (for Series~C
and D), i.e., making the scale separation poorer, results in a weak
decline of $\lambda$.
Theoretically, we would expect the turbulent decay rate to be
$\lambda=\chit k^2$, where $\chit=\chitz\equiv\urms/3\kf$ is the nominal
turbulent diffusivity in the case of perfect scale separation.
For poor scale separation, however, we expect
$\chit=\chitz/[1+(k/\kf)^2]$.
We see from the short lines overplotted on the left $y$ axis of
\Fig{pdecay_comp} that the actual decay rates are somewhat larger.

\begin{figure}[t!]
\includegraphics[width=\columnwidth]{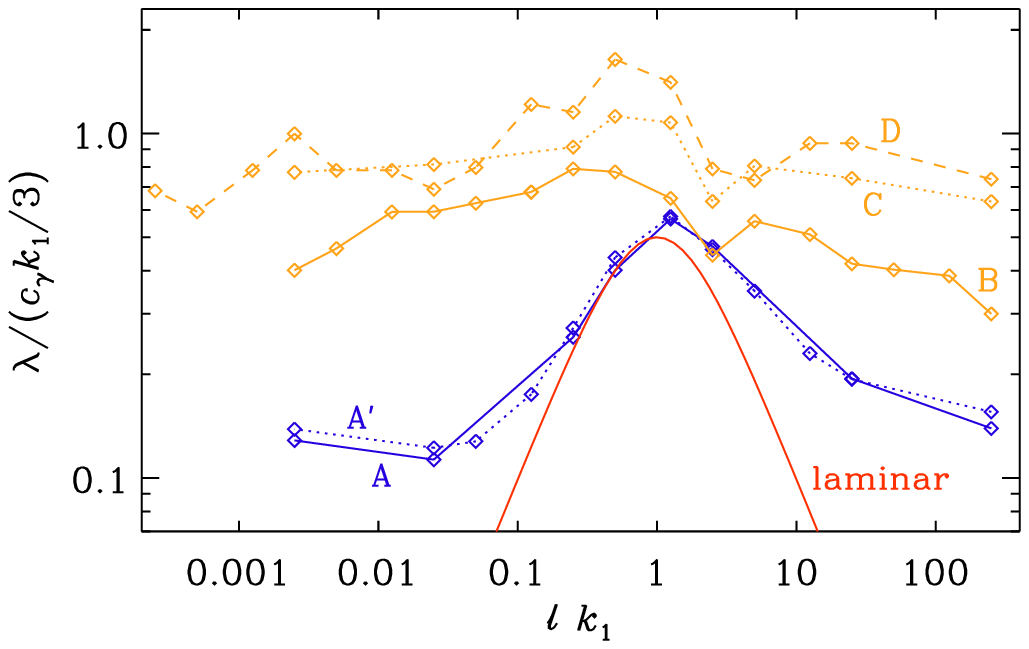}
\caption{
Similar to \Fig{pdecay_comp}, but the abscissa is scaled with $k_1$
instead of $k$, and the ordinate is divided by $k_1$ instead of $k$.  
Again, Series~A' and A are denoted by dotted and solid blue lines,
respectively, and Series~B, C, and D, are denoted by solid, dotted,
and dashed orange lines, respectively.
The curve for the laminar case is shown again in red.
\label{pdecay_comp_rescaled}}
\end{figure}

We note that in \Fig{pdecay_comp}, the decay rates of lines having
different values of $k$ (but the same value of $\kf$) are all separated
by factors that are close to $k$ itself.
To demonstrate that this is mostly the result of normalizing
$\lambda/\cgam$ by $k$, we show in \Fig{pdecay_comp_rescaled} the result
of normalizing $\lambda/\cgam$ by $k_1$, which is the same for all runs.
The lines are now no longer so strongly separated for different values
of $k$.
Note that the abscissa is also scaled by $k_1$ instead of $k$.
Consequently, the small peaks in $\lambda$ values near $k_1\ell=1$
occur at similar positions.

\begin{figure}[t!]
\includegraphics[width=\columnwidth]{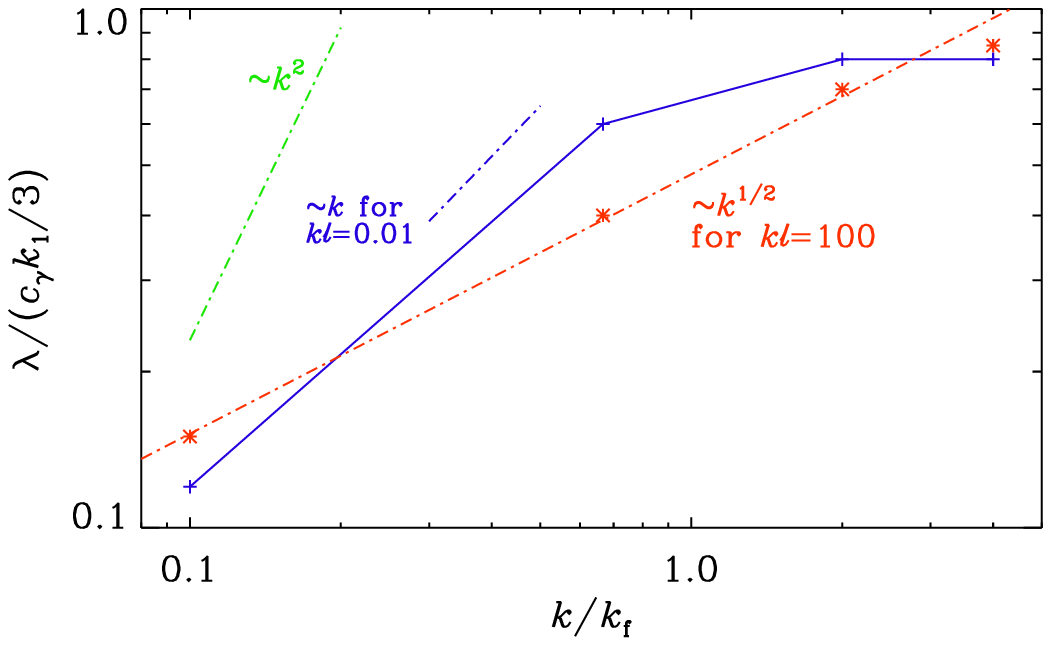}
\caption{
Dependence of $\lambda/(\cgam k_1/3)$ on $k/\kf$ for
$k\ell=0.01$ (blue, optically thick), and
$k\ell=100$ (red, optically thin).
The later obeys a $k^{1/2}$ scaling.
For comparison, we also show the linear scaling in $k$ (blue)
and the quadratic scaling (green).
\label{presults}}
\end{figure}

\subsection{Scale dependence}
\label{ScaleDependence}

The standard concept of turbulent diffusion with a diffusion
operator of the form $\chit\nabla^2$ requires one to have
sufficient scale separation, as is the case for our runs of
Series~A.
If scale separation is poor, the operator $\chit\nabla^2$
has to be replaced by a convolution in real space \cite{BRS08}.
This subject continues to attract significant attention,
especially in plasma physics \cite{BC20} and astrophysics
\cite{GE20,BenS21}.

In \Fig{presults} we summarize the results for $\lambda/(\cgam k_1/3)$
as a function of $k/\kf$ for $k\ell=0.01$ (optically thick regime),
and $k\ell=100$ (optically thin regime).
Neither of the two regimes exhibits a $k^2$ dependence, as would
be expected for a turbulent diffusion process with $k/\kf\ll1$, i.e.,
when the turbulent diffusivity is approximately scale-independent.
In the optically thick case, $\lambda$ increases approximately linearly
with $k$ for small values of $k$ and then reaches a maximum.
In the optically thin case, on the other hand, $\lambda$ increases
with $k$ approximately like $k^{1/2}$.

\begin{figure}[t!]
\includegraphics[width=\columnwidth]{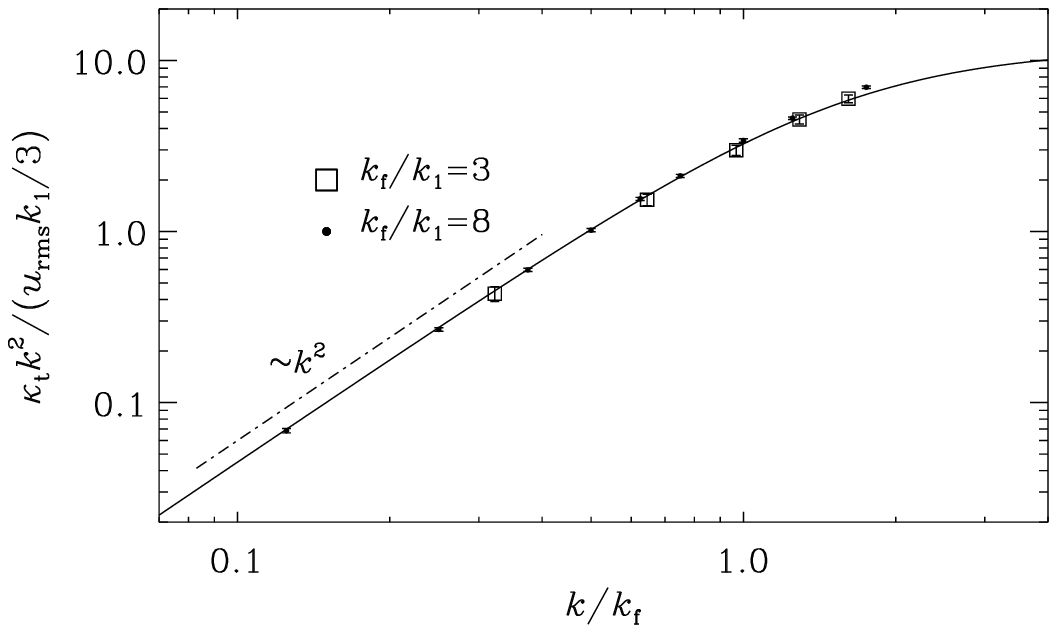}
\caption{
Dependence of the passive scalar diffusion rate $\kappa_{\rm t}(k) k^2$
on $k/\kf$ for scale separations $\kf/k_1=3$ and 8.
\label{p32d_mixing}}
\end{figure}

\subsection{Analogy with passive scalar diffusion}
\label{PassiveScalar}

In the optically thick regime, $k\ell\ll1$, where the radiative diffusion
approximation should be applicable, we might expect some analogy between
turbulent diffusion of active scalars (such as temperature) and
passive scalars (such as chemical concentrations).
For the latter, the scale dependence has previously been investigated
\cite{BSV09}, and it was found to be similar to that for magnetic fields
\cite{BRS08} in that both had a Lorentzian shape.
In these papers, the microphysical and turbulent diffusivities
were referred to as $\kappa$ and $\kappat$ for the passive scalar \cite{BSV09}
(not to be confused with the opacity in the present paper) and
$\eta$ and $\etat$ for the magnetic diffusivity \cite{BRS08}.
They have the same meaning as $\chi$ and $\chit$ in the present paper.
In these cases, we plot the time scale on which a large-scale
sinusoidal profile of the passive scalar or the magnetic field
gets diffused away.

We reproduce in \Fig{p32d_mixing} the result from
the test-field method for passive scalars \cite{BSV09}.
These authors also studied the effects of rotation and magnetic fields,
but those results were not used for the present comparison.
Their passive scalar diffusivity $\kappa_{\rm t}$ obeyed a Lorentzian
fit such that
\EQ
\kappa_{\rm t}(k)={\urms/3\kf\over1+(ak/\kf)^2},
\EN
where $a=0.62$ is an empirical parameter.
The corresponding decay rate, $\kappa_{\rm t}k^2$, is normalized
by $\urms k_1/3$ and shows a clear quadratic growth for small $k$
and levels off near $k=\kf$, as expected.

Similar Lorentzian fits have been found over a broad range of different
applications to turbulent magnetic diffusion: values of $a \approx 0.5$,
$a \approx 0.7$, and $a \approx 0.2$ were found for isotropic turbulence
\cite{BRS08}, anisotropic turbulence with shear \cite{MKTB09}, and
passive scalar diffusivity with shear \cite{MB10}, respectively.

\section{Conclusions}
\label{Conclusions}

Our work has demonstrated that the concept of turbulent diffusion carries
over to radiative turbulent diffusion as well, in both optically thick and
thin limits.
While this was expected for the optical thick limit, it was not obvious
how this would be modified in the optically thin limit, which is not a
diffusion process.
Instead, the optically thin case is characterized by Newtonian cooling,
which then turns into turbulent Newtonian cooling.
Both processes are shown to be scale-dependent, i.e., they are really
described by integral kernels.

We can now also answer the question regarding the combined effect of
decreased turbulence and small optical depth on the cooling at small
length scales.
As we have seen, turbulence always enhances the microphysical cooling
rates.
Thus, at small length scales where radiative diffusion is replaced by
the much less efficient Newtonian cooling, turbulence speeds up this
effect again.
Mathematically, this process is still treated like Newtonian cooling,
but now with a cooling time that is no longer given by $\ell/\cgam$,
but by the turbulent turnover time $(\urms\kf)^{-1}$.
Radiation no longer enters explicitly, except through the condition
$k\ell>1$ for turbulent Newtonian cooling, as opposed to $k\ell<1$ for
turbulent radiative diffusion.

As for the scope of future work, independent verifications of our results
would certainly be desirable.
In particular, it is conceivable that one can develop a test-field
method similar to that employed for passive scalars \cite{BSV09}.
It would also be useful to study the effects of turbulent radiative
diffusion and turbulent Newtonian cooling by comparing direct
numerical simulations with mean-field models.
This could be particularly insightful in more realistic situations
involving stratification, turbulence, and magnetic fields, which could
give rise to interesting phenomena such as magnetic spot formation
\cite{BRK16,PB18}.

\begin{acknowledgments}
We thank Matthias Rheinhardt for useful comments on the manuscript.
This work was supported in part by the Swedish Research Council,
grant 2019-04234.
We acknowledge the allocation of computing resources provided by the
Swedish National Allocations Committee at the Center for Parallel
Computers at the Royal Institute of Technology in Stockholm.
\end{acknowledgments}

\subsection*{Conflict of interest}

The authors have no conflicts to disclose.

\subsection*{Data Availability Statement}

The source code used for the simulations in this study, the {\sc Pencil Code}
\cite{PC}, is freely available on \url{https://github.com/pencil-code/}.
The DOI of the code is https://doi.org/10.5281/zenodo.2315093.
The simulation setup and the corresponding data \cite{TurbNewtonianCooling}
are freely available on \url{https://doi.org/10.5281/zenodo.4085411}.


\end{document}